\documentclass[singlecolumn,superscriptaddress,preprintnumbers,nofootinbib]{revtex4}
\usepackage{amsmath,amssymb,longtable}
\usepackage{graphicx}
\usepackage{epsfig,color}
\usepackage{ulem}
\usepackage{bm}
\usepackage{verbatim}
\newcommand{\footnoteref}[1]{\textsuperscript{\ref{#1}}}

\begin{document}

\title{Quark structure of the $\chi_{\rm c}(3P)$ and $X(4274)$ resonances and their strong and radiative decays}

\author{J. Ferretti\footnote{Corresponding author\label{ca}}}\email[]{jacopo.j.ferretti@jyu.fi}
\affiliation{Center for Theoretical Physics, Sloane Physics Laboratory, Yale University, New Haven, Connecticut 06520-8120, USA}
\affiliation{Department of Physics, University of Jyv\"askyl\"a, P.O. Box 35 (YFL), 40014 Jyv\"askyl\"a, Finland}
\author{E. Santopinto}\email[]{santopinto@ge.infn.it}
\affiliation{INFN, Sezione di Genova, Via Dodecaneso 33, 16146 Genova, Italy}
\author{M. Naeem Anwar\footnoteref{ca}}\email[]{m.anwar@fz-juelich.de}
\affiliation{Institut f\"ur Kernphysik, J\"ulich Center for Hadron Physics and Institute for Advanced Simulation, Forschungszentrum J\"ulich, 52425 J\"ulich, Germany}
\author{Yu Lu}\email[]{luyu@hiskp.uni-bonn.de}
\affiliation{Helmholtz-Institut f\"ur Strahlen- und Kernphysik and Bethe Center for Theoretical Physics, Universit\"at Bonn, D-53115 Bonn, Germany}

\begin{abstract}
We calculate the masses of $\chi_{\rm c}(3P)$ states with threshold corrections in a coupled-channel model. The model was recently applied to the description of the properties of $\chi_{\rm c}(2P)$ and $\chi_{\rm b}(3P)$ multiplets [Phys.\ Lett.\ B {\bf 789}, 550 (2019)].
We also compute the open-charm strong decay widths of the $\chi_{\rm c}(3P)$ states and their radiative transitions. 
According to our predictions, the $\chi_{\rm c}(3P)$ states should be dominated by the charmonium core, but they may also show small meson-meson components. The $X(4274)$ is interpreted as a $c \bar c$ $\chi_{\rm c1}(3P)$ state.
More informations on the other members of the $\chi_{\rm c}(3P)$ multiplet, as well as a more rigorous analysis of the $X(4274)$'s decay modes, are needed to provide further indications on the quark structure of the previous resonance.
\end{abstract}

\maketitle

\section{Introduction}
Recently, several new meson resonances have been discovered \cite{Tanabashi:2018oca,Esposito:2016noz,Olsen:2017bmm,Guo:2017jvc,Liu:2019zoy}. 
A fraction of them, the so-called $XYZ$ states, cannot be interpreted in terms of standard quark-antiquark degrees of freedom. 
Their description needs the introduction of more complicated exotic or multiquark structures.
A well-known example is the $X(3872)$ [now $\chi_{\rm c1}(3872)$] \cite{Choi:2003ue,Acosta:2003zx,Abazov:2004kp}. 
A wide range of theoretical descriptions of $XYZ$ states is available. These interpretations include: a) the compact tetraquark (or diquark-antidiquark) model \cite{Jaffe:1976ih,Barbour:1979qi,Weinstein:1983gd,SilvestreBrac:1993ss,Brink:1998as,Maiani:2004vq,Barnea:2006sd,Santopinto:2006my,Ebert:2008wm,Deng:2014gqa,Zhao:2014qva,Lu:2016cwr,Anwar:2017toa,Anwar:2018sol,Bedolla:2019zwg,Yang:2019itm,Ferretti:2020ewe}; b) the meson-meson molecular model \cite{Weinstein:1990gu,Manohar:1992nd,Tornqvist:1993ng,Martins:1994hd,Hanhart:2007yq,Thomas:2008ja,Baru:2011rs,Valderrama:2012jv,Aceti:2012cb,Guo:2013sya}; c) the interpretation in terms of kinematic or threshold effects caused by virtual particles \cite{Heikkila:1983wd,Lu:2017hma,Pennington:2007xr,Li:2009ad,Danilkin:2010cc,Ortega:2012rs,Ferretti:2013faa,Ferretti:2013vua,Achasov:2015oia,Kang:2016jxw,Lu:2016mbb,Lu:2017yhl,Ferretti:2018tco,Anwar:2018yqm}. 
Calculations of meson observables (like the spectrum or the decay widths) within the above pictures, when compared with the experimental data \cite{Tanabashi:2018oca}, may help to better understand the quark structure of $XYZ$ mesons.

In a previous paper \cite{Ferretti:2018tco}, we discussed a novel coupled-channel model approach to the spectroscopy and structure of heavy quarkonium-like mesons based on the Unquenched Quark Model (UQM) formalism \cite{Heikkila:1983wd,Ferretti:2013faa,Ferretti:2013vua,Ferretti:2014xqa,bottomonium,Bijker:2009up,Lu:2016mbb,Lu:2017hma,Lu:2017yhl,Anwar:2018yqm}.
In the UQM, quarkonium-like exotics are interpreted as the superposition of a heavy quarkonium core plus meson-meson molecular-type components.
In the approach of Ref.~\cite{Ferretti:2018tco}, the UQM formalism was used to compute the self-energy corrections to the bare masses of $\chi_{\rm c}(2P)$ and $\chi_{\rm b}(3P)$ states due to virtual particle effects.
However, differently from previous UQM calculations, see e.g. Refs.~\cite{Pennington:2007xr,Ferretti:2013faa,Ferretti:2013vua,Lu:2016mbb}, we did not perform a global fit to the whole heavy quarkonium spectrum. We applied the formalism to a single heavy quarkonium multiplet at a time. 
Moreover, we introduced a ``renormalization'' prescription for the UQM results.

Here, we make use of the same approach as Ref.~\cite{Ferretti:2018tco} to study the quark structure of the $X(4274)$ and $\chi_{\rm c}(3P)$ states by calculating their masses with threshold corrections. We also compute their open-charm strong decay widths in the $^3P_0$ pair-creation model \cite{Micu,LeYaouanc,Roberts:1992,Blundell:1995ev,Ackleh:1996yt,Barnes:2005pb,Godfrey:2015dva,Strong2015,Ferretti:2015rsa} and their radiative transitions in the UQM \cite{Anwar:2018yqm} formalism.
The $X(4274)$ [also known as $\chi_{\rm c1}(4274)$] was discovered by LHCb in the amplitude analysis of $B^+ \rightarrow J/\psi \phi K^+$ decays \cite{Aaij:2016iza}, even though a $3.1 \sigma$ evidence for a relatively narrow $J/\psi \phi$ mass peak near $4274\pm8$~MeV had been previously presented by CDF \cite{Aaltonen:2011at}.
Its quantum numbers are $I^G (J^{PC}) = 0^+ (1^{++})$ and its total decay width is $49\pm12$~MeV \cite{Tanabashi:2018oca}.

According to our coupled-channel model results, threshold effects should be small to medium-sized in the $\chi_{\rm c}(3P)$ multiplet. 
Our $^3P_0$ model prediction for the open-charm strong decay width of $X(4274)$ is compatible with the experimental data within the experimental error.
Therefore, it is reasonable to treat the $X(4274)$ as a charmonium state. However, due to the total lack of experimental data on the other members of the multiplet, we cannot exclude the presence of small meson-meson components in the $X(4274)$ wave function.
Our results for the radiative transitions of the $X(4274)$ and $\chi_{\rm c}(3P)$s will be an important check and may help to assess the quark structure of the previous resonances.

\section{Formalism}

\subsection{$^3P_0$ pair-creation model}
\label{3P0-formulas}
\begin{figure}[htbp]
\begin{center}
\includegraphics[width=7cm]{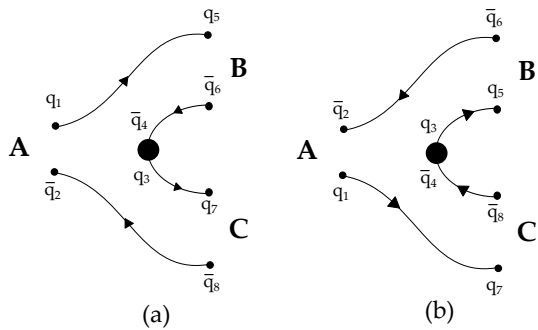}
\end{center}
\caption{Diagrams contributing to the $A\rightarrow BC$ decay process. $q_i$, with $i=1, ..., 4$, and $\bar{q}_j$, with $j=5, ..., 8$, are the quarks and antiquarks in the initial and final states, respectively. Picture from Ref.~\cite{bottomonium}. APS copyright.} 
\label{fig:diagrammi3P0}
\end{figure}

In the $^3P_0$ pair-creation model, the open-flavor strong decay of a hadron $A$ into hadrons $B$ and $C$ takes place in the rest frame of $A$. The decay proceeds via the creation of an additional $q \bar q$ pair with $J^{PC} = 0^{++}$ quantum numbers from QCD vacuum \cite{Micu,LeYaouanc,Roberts:1992} (see Fig. \ref{fig:diagrammi3P0}) and the width is computed as \cite{Micu,LeYaouanc,Ackleh:1996yt} 
\begin{equation}
	\label{eqn:3P0-decays-ABC}
	\Gamma_{A \rightarrow BC} = \Phi_{A \rightarrow BC}(q_0) \sum_{\ell} 
	\left| \left\langle BC q_0  \, \ell J \right| T^\dag \left| A \right\rangle \right|^2 \mbox{ }.
\end{equation}
where $\ell$ is the relative angular momentum between $B$ and $C$ and $J$ represents their total angular momentum.
The coefficient  
\begin{equation}
	\label{eqn:rel-PSF}
	\Phi_{A \rightarrow BC}(q_0) = 2 \pi q_0 \frac{E_B(q_0) E_C(q_0)}{M_A}  
\end{equation}
is the phase-space factor for the decay; it depends on the relative momentum $q_0$ between $B$ and $C$, the energies of the two decay products, $E_{B,C}(q_0)$, and the mass of the decaying meson, $M_A$.
We assume harmonic oscillator wave functions for the hadrons $A$, $B$ and $C$, depending on a single oscillator parameter $\alpha_{\rm ho}$; see \cite[Table II]{Ferretti:2013faa} and \cite[Table II]{Ferretti:2015rsa}. The values of the oscillator parameter, $\alpha_{\rm ho}$, and of the other pair-creation model parameters, $r_{\rm q}$ and $\gamma_0$, were fitted to the open-charm strong decays of higher charmonia \cite{Ferretti:2013faa}. 

Following Refs.~\cite{Ferretti:2013faa,Ferretti:2013vua,Ferretti:2014xqa,bottomonium,Ferretti:2015rsa}, we introduce a few changes in the $^3P_0$ pair-creation model operator, $T^\dag$.
These modifications include the substitution of the pair-creation strength, $\gamma_0$, with an effective one, $\gamma_0^{\rm eff}$, to suppress heavy quark pair-creation, see \cite[Eq. (12)]{bottomonium} and \cite{Kalashnikova:2005ui}, and the introduction of a Gaussian quark form-factor, because the pair of created quarks has an effective size \cite{Ferretti:2013faa,Ferretti:2013vua,Ferretti:2014xqa,bottomonium,Bijker:2009up}.

\subsection{Threshold mass-shifts in a coupled-channel model}
\label{Threshold mass-shifts}
We briefly summarize the main features of the coupled-channel model of Ref.~\cite{Ferretti:2018tco}. 
There, higher Fock components, $\left| BC \right\rangle$, due to virtual particle effects are superimposed on the $Q \bar Q$ bare meson wave functions, $\left| A \right\rangle$, of heavy quarkonium states.
One has \cite{Heikkila:1983wd,Ferretti:2013faa,Ferretti:2013vua,Ferretti:2014xqa,bottomonium,Bijker:2009up}:
\begin{equation}	
	\footnotesize
	\label{eqn:Psi-A}
	\begin{array}{l}	
	\left| \psi_A \right\rangle = {\cal N} \left[ \left| A \right\rangle + \displaystyle \sum_{BC \ell J} \int q^2 dq \, 
	\left| BC q \, \ell J \right\rangle \frac{ \left\langle BC q \, \ell J \right| T^{\dagger} \left| A \right\rangle}{M_A - E_B - E_C} \right] ~. 
	\end{array}
\end{equation}
The sum is extended over a complete set of meson-meson intermediate states $\left| BC \right\rangle$, with energies $E_{B,C}(q) = \sqrt{M_{B,C}^2 + q^2}$; $M_A$ is the physical mass of the meson $A$; $q$ is the on-shell momentum between $B$ and $C$, $\ell$ is the relative orbital angular momentum between them, and $J$ is the total angular momentum, with ${\bf J} = {\bf J}_B + {\bf J}_C + {\bm\ell}$.
Finally, the amplitudes $\left\langle BC q \, \ell J \right| T^{\dagger} \left| A \right\rangle$ are computed within the $^3P_0$ pair-creation model of Sec. \ref{3P0-formulas}.

In the coupled-channel approach of Ref.~\cite{Ferretti:2018tco}, one can study a single multiplet at a time, like $\chi_{\rm c}(2P)$ or $\chi_{\rm b}(3P)$.
The physical masses of the meson multiplet members are given by
\begin{equation}
	\label{eqn:new-Ma}
	M_A = E_A + \Sigma(M_A) + \Delta_{\rm th} \mbox{ }.
\end{equation}
In the previous equation, 
\begin{equation}
	\label{eqn:self-a}
	\Sigma(M_A) = \sum_{BC} \int_0^{\infty} q^2 dq \mbox{ } 
	\frac{\left| \left\langle BC q  \, \ell J \right| T^\dag \left| A \right\rangle \right|^2}{M_A - E_B(q) - E_C(q)}  
\end{equation}
is a self-energy correction, $E_A$ is the bare mass of the meson $A$, and $\Delta_{\rm th}$ is a free parameter.
Contrary to our previous UQM studies \cite{Ferretti:2013faa,Ferretti:2013vua}, the bare meson masses $E_A$ are not fitted to the whole charmonium spectrum. Their values are directly extracted from the relativized QM predictions of Refs.~\cite{Barnes:2005pb,Godfrey:1985xj}.
The UQM model parameters, which we need in the calculation of the $\left\langle BC q  \, \ell J \right| T^\dag \left| A \right\rangle$ vertices, were fitted to the open-flavor strong decays of charmonia; see \cite[Table II]{Ferretti:2013faa} and \cite[Table II]{Ferretti:2015rsa}.
Thus, for each multiplet $\Delta_{\rm th}$ is the only free parameter. It is defined as the smallest self-energy correction (in terms of absolute value) among those of the multiplet members; see Sec. \ref{chiC(3P)} and \cite[Secs. 2.2 and 2.3]{Ferretti:2018tco}.
The introduction of $\Delta_{\rm th}$ in Eq. (\ref{eqn:new-Ma}) represents our ``renormalization'' or ``subtraction'' prescription for the threshold mass-shifts in the UQM.

\subsection{Radiative Transitions in the QM and UQM formalisms}
\label{trans}
Radiative transitions of higher charmonia are of considerable interest, since they can shed light on the structure of $c \bar c$ states and provide one of the few pathways between different $c\bar{c}$ multiplets. Particularly, for those states which cannot be directly produced at $e^+ e^-$ colliders (such as $P$-wave charmonia), the radiative transitions serve as an elegant probe to explore such systems.

In the quark model, the electric dipole ($E1$) transitions can be expressed as~\cite{Eichten:1978tg,Gupta,Kwong:1988ae}
\begin{equation}
	\label{eqn:E1X3872}
	\Gamma_{E1, {\rm QM}} = \frac{4\alpha e_c^2}{3} ~ C_{AB}   
	\left| \left\langle R_B \right| r \left| R_A \right\rangle \right|^2 \frac{E_\gamma^3 \tilde E_B}{M_A} 
	~ \delta_{S_{\rm A},S_{\rm B}}  \mbox{ }.
\end{equation}
Here, $e_c = \frac{2}{3}$ is the $c$-quark charge, $\alpha$ the fine structure constant, $ E_\gamma$ denotes the energy of the emitted photon, and $\tilde E_B = \sqrt{M_B^2 + E_\gamma^2}$ is the total energy of the final meson. The spatial matrix elements 
\begin{equation}
	\label{eqn:overlap}
	\left\langle R_B \right| r \left| R_A \right\rangle = \int_0^\infty r^3 dr \mbox{ } R^*_{n_B ,L_B}(r) \mbox{ } 
	R_{n_A, L_A}(r)
\end{equation}
involve the initial and final meson radial wave functions which we obtained numerically; for more details, see Refs.~\cite{Lu:2017yhl,Anwar:2018yqm}.
The angular matrix elements $C_{AB}$ are given by
\begin{equation}
	C_{AB} = \mbox{max}(L_A,L_B) ~ (2J_B+1) \left\{ \begin{array}{rcl} L_B & J_B & S_A \\ 
	J_A & L_A & 1\end{array} \right\}^2 \mbox{ },
\end{equation}
where $S_{A,B}$, $L_{A,B}$ and $J_{A,B}$ are the spin, orbital angular momentum and total angular momentum of the initial/final charmonia, respectively. 

In the UQM formalism, the wave function of a heavy quarkonium state consists of both a $Q \bar Q$ valence configuration and meson-meson higher Fock components, which are the result of the creation of light $q \bar q$ pairs from the vacuum.
Therefore, the heavy quarkonium bare meson wave function has to be properly renormalized. The probability of the charmonium core of a meson $A$ can be computed using the following relation 
\begin{equation}	
	\label{eqn:Psi-A}
	\begin{array}{l}	
	P_{c\bar{c}} (A)= \left[ 1 + \displaystyle \sum_{BC \ell J} \int q^2 dq \, 
        \frac{ \left| \left\langle BC q \, \ell J \right| T^{\dagger} \left| A \right\rangle \right|^2}{(M_A - E_B - E_C)^2} \right]^{-1}~. 
	\end{array}
\end{equation}
In our specific case, the radial wave functions $R_{A,B}$ of Eq. (\ref{eqn:overlap}) have to be multiplied by the factors $P_{c \bar c}(A,B) \le 1$, which are the probabilities of finding the wave functions of the $A$ and $B$ states in their valence components.
Given this, in the UQM formalism the radiative decay width becomes
\begin{equation}
	\label{eqn:E1-UQM}
	\Gamma_{E1, {\rm UQM}} = \Gamma_{E1, {\rm QM}} P_{c \bar c}(A) P_{c \bar c}(B)  \mbox{ }.
\end{equation}

\section{Results}
\subsection{Open-charm strong decays of $\chi_{\rm c}(3P)$ states}
\label{Open-charm strong decays}
In this section, we calculate the open-charm strong decay widths of $\chi_{\rm c}(3P)$ states within the $^3P_0$ pair-creation model. The  main features of the model are briefly described in Sec. \ref{3P0-formulas}.
When available, we extract the masses of both the initial- and final-state mesons from the PDG \cite{Tanabashi:2018oca}; otherwise, we use the relativized QM predictions of Refs.~\cite{Barnes:2005pb,Godfrey:1985xj}.
Our theoretical results are given in Table \ref{tab:ChiC(3P)-strong} and can be compared to the $^3P_0$ pair-creation model results of \cite[Table XI]{Barnes:2005pb}.
\begin{table}
\centering
\begin{ruledtabular}
\begin{tabular}{ccc|ccc} 
State                        & Channel           & Width           & State                        & Channel           & Width \\
                                &                          &           [MeV] &                           &     &      [MeV]  \\                                
\hline
$\chi_{\rm c1}(3P)$ & $D \bar D^*$                             & 6.6   & $\chi_{\rm c0}(3P)$ & $D \bar D$                                & 4.0    \\
                                & $D^* \bar D^*$                          & 28.0 &                                 & $D^* \bar D^*$                          & 35.0 \\
                                & $D \bar D_0^*$                         & 0.2   &                                 & $D \bar D_1(2420)$                  & 3.4  \\
                                & $D_{\rm s} \bar D_{\rm s}^*$    &  6.3  &                                 & $D \bar D_1(2430)$                  & 0.8    \\                                
                                & $D_{\rm s}^* \bar D_{\rm s}^*$ &  2.5  & & $D_{\rm s} \bar D_{\rm s}$       & 2.6    \\    
                                &                                                  &         & & $D_{\rm s}^* \bar D_{\rm s}^*$ & 6.6    \\                                   
\hline
$h_{\rm c}(3P)$       & $D \bar D^*$                             & 4.3   & $\chi_{\rm c2}(3P)$ & $D \bar D$                                & 9.3   \\ 
                                & $D^* \bar D^*$                          &19.9 &                                 & $D \bar D^*$                            & 5.8    \\
                                & $D \bar D_0^*$                         & 4.3   &                                 & $D^* \bar D^*$                         & 23.7  \\
                                & $D \bar D_1(2420)$                  & 0.03 &                               & $D \bar D_1(2420)$                 & 6.0 \\   
                                & $D \bar D_1(2430)$                  & 0.1   &                                 & $D \bar D_1(2430)$                & 3.2  \\ 
                                & $D_{\rm s} \bar D_{\rm s}^*$    & 7.9   &                                 & $D \bar D_2^*(2460)$              & 1.4  \\                                 
                                & $D_{\rm s}^* \bar D_{\rm s}^*$ & 5.1   &                                 & $D_{\rm s} \bar D_{\rm s}$       & 0.03 \\                                                                                                                                                                                                                                                                                                                                                                 
                                &                                                  &         &                                 & $D_{\rm s} \bar D_{\rm s}^*$    & 4.3 \\                               
                                &                                                  &         &                                 & $D_{\rm s}^* \bar D_{\rm s}^*$ & 4.7 \\
\end{tabular}
\end{ruledtabular}
\caption{Open-charm strong decays of $\chi_{\rm c}(3P)$ states in the $^3P_0$ pair-creation model. The values of the $\chi_{\rm c}(3P)$ masses are taken from Refs.~\cite{Barnes:2005pb,Godfrey:1985xj} (see also Table \ref{tab:ChiC(3P)-splittings}, second column), except for the value of the $\chi_{\rm c1}(3P)$ [or $X(4274)$] mass, which is extracted from the PDG \cite{Tanabashi:2018oca}. The values of the charmed and charmed-strange meson masses are taken from the PDG \cite{Tanabashi:2018oca}, the mixing angle between $D_1(1P_1)$ and $D_1(1P_1')$ states is taken from \cite[Table III]{Ferretti:2015rsa}.}
\label{tab:ChiC(3P)-strong}  
\end{table}

It is worth noting that: I) our predictions are of the same order of magnitude as those of \cite[Table XI]{Barnes:2005pb}. The discrepancies are in the order of $10-20\%$, except for the $h_{\rm c}(3P)$, where they are larger. These differences between our results and those of Ref.~\cite{Barnes:2005pb} arise partly because of different choices of the $^3P_0$ model parameters, and partly because of the values of the masses of the decaying mesons given as inputs in the calculations. In particular, in our case we use for the decaying meson masses either the experimental values \cite{Tanabashi:2018oca} or relativized QM predictions \cite{Godfrey:1985xj}. On the contrary, in Ref.~\cite{Barnes:2005pb} the authors extracted the masses from a non-relativistic potential model fit to the charmonium spectrum. Moreover, the use of different masses for the $c \bar c$ decaying mesons determines the opening of decay channels, like $\chi_{\rm c2}(3P) \rightarrow D \bar D_2^*(2460)$, which were below threshold in \cite[Table XI]{Barnes:2005pb}. Finally, as a check we have also computed the decay widths of $\chi_{\rm c}(3P)$s by using the same input masses and model parameters as Ref. \cite{Barnes:2005pb} and we have obtained the same results therein; II) according to our results, the $\chi_{\rm c}(3P)$s are characterized by relatively large open-charm widths, which are of the order of $40-60$~MeV. If our predictions were confirmed by the experiments, we may argue that $\chi_{\rm c}(3P)$ mesons are charmonium-like states, with their wave functions being dominated by a $c \bar c$ core; III) of particular interest are our results for the $\chi_{\rm c1}(3P)$ state. Specifically, our theoretical prediction for the total open-charm width of the $\chi_{\rm c1}(3P)$, i.e. $43.6$~MeV, is compatible with the total experimental width of the $\chi_{\rm c1}(4274)$ \cite{Tanabashi:2018oca}, namely $49\pm12$~MeV, under the hypothesis that the open-charm contribution to the total width of the $\chi_{\rm c1}(4274)$ is the dominant one.
As discussed in the previous point, this suggests that the wave function of the $\chi_{\rm c1}(4274)$ should be dominated by the charmonium component; 
IV) the results of Table \ref{tab:ChiC(3P)-strong} are obtained by using the $^3P_0$ pair-creation model with Simple Harmonic Oscillator (SHO) wave functions for both the parent and daughter hadrons \cite{Micu,LeYaouanc,Roberts:1992,Anwar:2016mxo}.
The $^3P_0$ model is quite sensitive to the form of the wave functions used.
This is why several studies have discussed the use of ``more realistic'' wave functions \cite{Kokoski:1985is,Capstick:1992th} or the limitations of the model \cite{Lu:2017hma,Kokoski:1985is,Capstick:2000qj}.
If we used different forms of meson wave functions, we would get slightly different predictions.
However, it is very interesting to observe that our result for the total open-flavor strong decay width of the $\chi_{\rm c1}(4274)$ is not only compatible with the experimental data, but it is also similar to that of Ref. \cite{Barnes:2005pb}, where the authors used different values of the $^3P_0$ model parameters and a slightly different value of the input mass of the $\chi_{\rm c1}(4274)$.
This is a further indication of the fact that the $\chi_{\rm c1}(4274)$ state should be dominated by the $c \bar c$ core component. 
We expect that the use of ``more realistic'' wave functions would not change the previous conclusion.

\subsection{$E1$ radiative transitions of $\chi_{\rm c}(3P)$ states}
\label{E1 charmonium}
Here, we discuss our UQM results for the $E1$ radiative transitions of $\chi_{\rm c}(3P)$ states.
Our predictions, denoted as $\Gamma_{E1, \rm UQM}$ and computed by means of Eq. (\ref{eqn:E1-UQM}), are given in Table \ref{tab:ChiC(3P)-E1}; see also Tables \ref{tab:others-E1} and \ref{tab:ChiC(3P)-E1-M1-exp}.
The QM widths of Eq. (\ref{eqn:E1X3872}), $\Gamma_{E1, \rm QM}$, are computed by using Cornell potential model \cite{Eichten:1978tg,Lu:2016mbb,Lu:2017yhl} wave functions for both the parent and daughter charmonium states.
Our results for the $\Gamma_{E1, \rm QM}$ widths coincide with those reported in Ref.~\cite{Barnes:2005pb}; therefore, they are not shown in the present paper.
\begin{table}[htbp]
\centering
\begin{ruledtabular}
\begin{tabular}{cccc} 
Transition & $P_{c\bar{c}}(A)$ & $P_{c\bar{c}}(B)$ & $\Gamma_{E1, \rm UQM}$ \\  
 &  &  & [keV] \\      
\hline                     
$\chi_{\rm c2}(3P) \rightarrow \psi(3S) + \gamma$ & 1.000 & 0.940 & 482 \\
$\chi_{\rm c1}(3P) \rightarrow \psi(3S) + \gamma$ & 1.000 & 0.940 & 287 \\
$\chi_{\rm c0}(3P)  \rightarrow \psi(3S) + \gamma$ & 0.960 & 0.940 & 99 \\
$h_{\rm c}(3P) \rightarrow \eta_c(3S) + \gamma$ & 1.000 & 0.930 & 258 \\
$\chi_{\rm c2}(3P) \rightarrow \psi(2S) + \gamma$ & 1.000 & 0.660 & 36 \\
$\chi_{\rm c1}(3P) \rightarrow \psi(2S) + \gamma$ & 1.000 & 0.660 & 29 \\
$\chi_{\rm c0}(3P) \rightarrow \psi(2S) + \gamma$ & 0.960 & 0.660 & 20 \\
$h_{\rm c}(3P) \rightarrow \eta_c(2S) + \gamma$ & 1.000 & 0.720 & 53 \\
$\chi_{\rm c2}(3P) \rightarrow J/\psi + \gamma$ & 1.000 & 0.770 & 26 \\
$\chi_{\rm c1}(3P) \rightarrow J/\psi + \gamma$ & 1.000 & 0.770 & 24 \\
$\chi_{\rm c0}(3P) \rightarrow J/\psi + \gamma$ & 0.960 & 0.770 & 19 \\
$h_{\rm c}(3P) \rightarrow \eta_c + \gamma$ & 1.000 & 0.800 & 57 \\
$\chi_{\rm c2}(3P) \rightarrow \psi_3(2^3{D}_3) + \gamma$ & 1.000 & 0.960 & 143 \\
$\chi_{\rm c2}(3P) \rightarrow \psi_2(2^3{D}_2) + \gamma$ & 1.000 & 0.970 & 30 \\
$\chi_{\rm c2}(3P) \rightarrow \psi_1(2^3{D}_1) + \gamma$ & 1.000 & 0.960 & 2 \\
$\chi_{\rm c1}(3P) \rightarrow \psi_2(2^3{D}_2) + \gamma$ & 1.000 & 0.970 & 56 \\
$\chi_{\rm c1}(3P) \rightarrow \psi_1(2^3{D}_1) + \gamma$ & 1.000 & 0.960 & 18 \\
$\chi_{\rm c0}(3P) \rightarrow \psi_1(2^3{D}_1) + \gamma$ & 0.960 & 0.960 & 4 \\
$h_{\rm c}(3P) \rightarrow \eta_{\rm c2}(2^1{D}_2) + \gamma$ & 1.000 & 0.970 & 96 \\
$\chi_{\rm c2}(3P) \rightarrow \psi_3(1^3{D}_3) + \gamma$ & 1.000 & 0.660 & 0 \\
$\chi_{\rm c2}(3P) \rightarrow \psi_2(1^3{D}_2) + \gamma$ & 1.000 & 0.570 & 0 \\
$\chi_{\rm c2}(3P) \rightarrow \psi_1(1^3{D}_1) + \gamma$ & 1.000 & 0.680 & 0 \\
$\chi_{\rm c1}(3P) \rightarrow \psi_2(1^3{D}_2) + \gamma$ & 1.000 & 0.570 & 0 \\
$\chi_{\rm c1}(3P) \rightarrow \psi_1(1^3{D}_1) + \gamma$ & 1.000 & 0.680 & 0 \\
$\chi_{\rm c0}(3P) \rightarrow \psi_1(1^3{D}_1) + \gamma$ & 0.960 & 0.680 & 0 \\
$h_{\rm c}(3P) \rightarrow \eta_{\rm c2}(1^1{D}_2) + \gamma$ & 1.000 & 0.590 & 0 \\
\end{tabular}
\end{ruledtabular}
\caption{$E1$ radiative decay amplitudes of $\chi_{\rm c}(3P)$ states in the UQM formalism. The second and third columns report the normalizations of initial ($A$) and final ($B$) charmonium. Our UQM predictions, $\Gamma_{E1, \rm UQM}$, are computed according to Eq. (\ref{eqn:E1-UQM}), where the QM results, $\Gamma_{E1, \rm QM}$, can be extracted from Ref. \cite{Barnes:2005pb}. The large values of the $P_{c\bar{c}}$ probabilities for $\chi_{\rm c}(3P)$s are due to the conservative renormalization prescription used in our UQM calculation of the radiative decays. See the explanation in the text and also in Ref. \cite{Anwar:2018yqm}.}
\label{tab:ChiC(3P)-E1}  
\end{table}
The UQM predictions, denoted as $\Gamma_{E1, \rm UQM}$, are calculated by renormalizing the $A$ and $B$ meson wave functions according to the valence probabilities $P_{c\bar{c}}(A)$ and $P_{c\bar{c}}(B)$. 

In this section, we make use of a renormalization prescription different from that of the UQM-based coupled-channel model of Ref. \cite{Ferretti:2018tco} and Secs. \ref{Threshold mass-shifts} and \ref{chiC(3P)}.
The reason behind this choice is the necessity of simplifying our calculations, in which a large amount of $A$ and $B$ states is taken into account ($1S$, $2S$, $1P$, $1D$, and so on).
In particular, the calculation of the probabilities $P_{c\bar{c}}(A,B)$ of Table \ref{tab:ChiC(3P)-E1} is performed in the standard UQM formalism \cite{bottomonium,Lu:2016mbb}, with the model parameter values, $\alpha_{\rm ho} = 0.5$ GeV and $\gamma_0 = 0.4$, extracted from Ref.~\cite{Barnes:2005pb}. 
The renormalization prescription we use here is the same as Ref. \cite{Anwar:2018yqm} and consists in: I) considering $1S1S$ intermediate states only, both in the case of the parent and daughter charmonia. This is the approximation used in the large majority of the UQM calculations for mesons; II) discarding the contributions of the open-channels to the wave function renormalization~\cite{Heikkila:1983wd}.
The latter assumption is used to deal with those $\chi_{\rm c}(3P)$ states which are above the $D^{(*)} \bar D^{(*)}$ and $D_{\rm s}^{(*)} \bar D_{\rm s}^{(*)}$ thresholds.
Recently, the above prescription was used to study the radiative decays of $\chi_{\rm b}(3P)$ bottomonia \cite{Anwar:2018yqm}, and a quite reasonable agreement with the recent CMS measurements was found.

It is worth noting that: I) the radiative decay widths of $\chi_{\rm c}(3P)$ states span a wide interval, from $\mathcal{O}(300 \mbox{ MeV})$ to $\mathcal{O}(1 \mbox{ MeV})$, in the case of $3P \rightarrow 3S + \gamma$ and $3P \rightarrow 1D + \gamma$ transitions, respectively. In particular, the $3P \rightarrow 3S + \gamma$ decay widths are quite large; thus, they might be observed in the next few years; II) our UQM results for $\chi_{\rm c}(3P)$ states are roughly of the same order of magnitude as the QM ones \cite{Barnes:2005pb}. The difference between them is of the order of $5-10\%$ in the case of the $3P \rightarrow 3S + \gamma$ transitions and $30-40\%$ for $3P \rightarrow 2S + \gamma$ decays. This is a confirmation of our statement that loop effects can play a relatively important role in determining the properties of $\chi_{\rm c}(3P)$s, though their importance is far from being conclusive.
In this respect, it is interesting to estimate the importance of loop effects in the case of other charmonium radiative transitions. See Appendix \ref{appendix A}, Tables \ref{tab:others-E1} and \ref{tab:ChiC(3P)-E1-M1-exp}, and the QM results of Ref. \cite{Barnes:2005pb}. For example, consider the $\chi_{\rm c2}(2P) \rightarrow \psi_3(1^3{\rm D}_1) + \gamma$ decay, where the ratio between the QM \cite{Barnes:2005pb} and UQM widths is almost a factor of 2.5; III) we also show that the $E1$ transition widths of $\chi_{\rm c}(3P)$s into $J/\psi+\gamma$ are one order of magnitude suppressed with respect to those into $\psi(3S)+\gamma$. A similar pattern was previously observed in the $\chi_{\rm b}(3P)$ case \cite{Anwar:2018yqm}; IV) finally, the UQM results depend on the specific renormalization prescription taken into account. Thus, the use of a different renormalization prescription will necessarily produce quite different results.

In conclusion, our UQM results may provide solid references to search for the other members of the $\chi_{\rm c}(3P)$ multiplet by analyzing the $\chi_{\rm c}(3P)\rightarrow \psi(2S,3S) + \gamma$ radiative transitions. 
Recently, the CMS Collaboration was able to distinguish for the first time between two candidates of the bottomonium $3P$ multiplet, $\chi_{b1}(3P)$ and $\chi_{b2}(3P)$, through their $\Upsilon(nS) + \gamma$ ($n=1,2,3$) decays \cite{Sirunyan:2018dff}.
We expect the charmonium $3P$ multiplet to be easily searched by means of the same strategy.

\subsection{Threshold mass shifts within the $\chi_{\rm c}(3P)$ multiplet}
\label{chiC(3P)}
We calculate the relative threshold mass shifts between the $\chi_{\rm c}(3P)$ multiplet members due to a complete set of $1S 1P$ meson-meson loops, like $D D_0^*(2300)$, $DD_1(2420)$, and so on.\footnote{It is worth noting that in the coupled-channel model calculation of the threshold corrections of $\chi_{\rm c}(2P)$ and $\chi_{\rm b}(3P)$ states of Ref.~\cite{Ferretti:2018tco}, only $1S 1S$ open-flavor loops were taken into account. Here, on the contrary, we only include $1S 1P$ loops. $1S 1S$ loops in the former case and $1S 1P$ loops in the latter are identified as the complete sets of intermediate states which are closer in energy to the multiplet members \cite[Sec. 2]{Ferretti:2018tco}. The other $nL$ $n' L'$ loops, with $n, n' = 1, 2, 3, 4, ...$ and $L, L' = S, P, D, F, ...$, are farther in energy and can be neglected.} 
As shown in Ref.~\cite{Ferretti:2013faa}, charmonium loops, like $\eta_{\rm c} \chi_{\rm c0}(1P)$, are negligible because of the suppression mechanism of \cite[Eq. (12)]{bottomonium}. Therefore, these loops are not taken into account in the calculation of the self-energy corrections of $\chi_{\rm c}(3P)$ states.
\begin{table}
\footnotesize
\centering
\begin{ruledtabular}
\begin{tabular}{ccccc} 
State                & $D D_0^*(2300)$ & $DD_1(2420)$  & $D D_1(2430)$ & $D D_2^*(2460)$ \\
\hline
$h_c(3P)$          & $-12.3$  & --  & $-0.5$  & $-30.6$  \\
$\chi_{c0}(3P)$  & --  & $-41.6$ & $-28.3$ & --  \\
$\chi_{c1}(3P)$  & $-0.2$  & $-13.3$ & $-12.9$ & $-18.1$  \\
$\chi_{c2}(3P)$  & --  & $-15.7$ & $-14.2$ & $-15.9$  \\
\hline
State                & $D^* D_0^*(2300)$ & $D^*D_1(2420)$  & $D^* D_1(2430)$ & $D^* D_2^*(2460)$ \\
\hline
$h_c(3P)$          & $-0.3$ & $-33.5$ & $-35.3$ & $-28.3$ \\
$\chi_{c0}(3P)$  & $-10.9$ & $-22.0$ & $-16.4$ & $-39.6$ \\
$\chi_{c1}(3P)$  & $-11.7$ & $-22.7$ & $-26.6$ & $-28.4$ \\
$\chi_{c2}(3P)$  & $-17.3$ & $-21.3$ & $-23.2$ & $-42.8$ \\
\hline
State                & $D_{\rm s} D_{\rm s0}^*(2317)$ & $D_{\rm s} D_{\rm s1}(2460)$  & $D_{\rm s} D_{\rm s1}(2536)$ & $D_{\rm s} D_{\rm s2}^*(2573)$ \\
\hline
$h_c(3P)$          & $-1.5$  & --   & $-0.1$  & $-3.6$  \\
$\chi_{c0}(3P)$  & --  & $-2.6$   & $-3.2$  & --  \\
$\chi_{c1}(3P)$  & $-0.1$  & $-1.7$   & $-0.8$  & $-2.3$  \\
$\chi_{c2}(3P)$  & --  & $-1.8$   & $-1.2$  & $-1.8$  \\
\hline
State                & $D_{\rm s}^* D_{\rm s0}^*(2317)$ & $D_{\rm s}^*D_{s1}(2460)$  & $D_{\rm s}^* D_{s1}(2536)$ & $D_{\rm s}^* D_{\rm s2}^*(2573)$ \\
\hline
$h_c(3P)$          & $-0.1$  & $-4.5$   & $-3.4$  & $-4.6$  \\
$\chi_{c0}(3P)$  & $-0.7$  & $-3.1$   & $-1.3$  & $-6.5$  \\
$\chi_{c1}(3P)$  & $-0.8$  & $-3.0$   & $-3.1$  & $-4.8$  \\
$\chi_{c2}(3P)$  & $-1.4$  & $-2.9$   & $-2.1$  & $-6.6$  \\
\hline
State                & & & & $\Sigma(M_A)$ \\
\hline
$h_c(3P)$          &   &    &   & $-159$  \\
$\chi_{c0}(3P)$  &   &    &   & $-176$  \\
$\chi_{c1}(3P)$  &   &    &   & $-151$  \\
$\chi_{c2}(3P)$  &   &    &   & $-168$  \\
\end{tabular}
\end{ruledtabular}
\caption{Self-energy corrections, $\Sigma(M_A)$ (in MeV), to the bare masses of $\chi_{\rm c}(3P)$ states, calculated via Eq. (\ref{eqn:self-a}). The values of the UQM parameters are extracted from \cite[Table II]{Ferretti:2013faa}.
The first rows show the partial contributions to $\Sigma(M_A)$ from channels $BC$, such as $D D_0^*(2300)$, $DD_1(2420)$, and so on. The last rows provide the total results, obtained by summing the previous partial contributions. The contributions of those channels denoted by -- are suppressed by selection rules.} 
\label{tab:Mass-shifts}  
\end{table}

Following Ref.~\cite{Ferretti:2018tco}, the values of the bare meson masses, $E_A$, are extracted from the relativized QM predictions of \cite[Table I, sixth column]{Barnes:2005pb} and \cite{Godfrey:1985xj}. 
We have: $E_{h_{\rm c}(3P)} = 4318$~MeV, $E_{\chi_{\rm c0}(3P)} = 4292$~MeV, $E_{\chi_{\rm c1}(3P)} = 4317$~MeV and $E_{\chi_{\rm c2}(3P)} = 4337$~MeV. 
The values of the physical masses, $M_A$, of the $\chi_{\rm c}(3P)$ states should be extracted from the data \cite{Tanabashi:2018oca}. 
However, except for the mass of the $\chi_{\rm c1}(4274)$, $4274^{+8}_{-6}$~MeV, there are no experimental results for the masses of the remaining and still unobserved $\chi_{\rm c}(3P)$ states, namely the $h_{\rm c}(3P)$, $\chi_{\rm c0}(3P)$ and $\chi_{\rm c2}(3P)$. Therefore, for the physical masses of the previous unobserved states we use the same values as the bare ones \cite{Barnes:2005pb,Godfrey:1985xj}.
Moreover, for simplicity, we do not consider mixing effects between $\left |1^1P_1 \right\rangle$ and $\left|1^3P_1 \right\rangle$ charmed and charmed-strange mesons in the self-energy calculation of this section. Therefore, for the wave functions of the previous states we make the assumptions: $\left |1P_1 \right\rangle \simeq \left |1^1P_1\right\rangle$ and $\left|1P_1' \right\rangle \simeq \left|1^3P_1 \right\rangle$.
The self-energy corrections are computed according to the UQM formalism of Sec. \ref{Threshold mass-shifts} and Refs.~\cite{Ferretti:2013faa,Ferretti:2013vua}. 
Our results are reported in Table \ref{tab:Mass-shifts}. 

Compared to our previous results for $\chi_{\rm c}(2P)$s and $\chi_{\rm b}(3P)$s \cite{Ferretti:2018tco}, the present results for $\chi_{\rm c}(3P)$ states are more model-dependent. The reason is the lack of experimental data for three of the four multiplet members.
Finally, our results for the ``renormalized'' threshold corrections, $\Sigma(M_A) - \Delta_{\rm th}$, and the calculated physical masses, $M_A^{\rm th}$, of $\chi_{\rm c}(3P)$ states are reported in Table \ref{tab:ChiC(3P)-splittings}. 
\begin{table}[htbp]
\centering
\begin{ruledtabular}
\begin{tabular}{ccccc} 
State                        & $E_A$             & $\Sigma(M_A) - \Delta_{\rm th}$           & $M_A^{\rm th}$           & $M_A^{\rm exp}$ \\
                                &             [MeV]  &                                       [MeV] &  [MeV]                        &  [MeV] \\
\hline
$h_{\rm c}(3P)$       & 4318               & $-8$                    & 4310          & -- \\
$\chi_{\rm c0}(3P)$ & 4292               & $-25$                   & 4267          & --  \\
$\chi_{\rm c1}(3P)$ & 4317               & 0                          & 4317          & $4274^{+8}_{-6}$  \\
$\chi_{\rm c2}(3P)$ & 4337               & $-17$                   & 4320          & --  \\                                                      
\end{tabular}
\end{ruledtabular}
\caption{Comparison between the experimental masses \cite{Tanabashi:2018oca} of $\chi_{\rm c}(3P)$ states and our theoretical predictions, as explained in the text. The bare mass values, $E_A$, are extracted from Refs.~\cite{Barnes:2005pb,Godfrey:1985xj}.}
\label{tab:ChiC(3P)-splittings}  
\end{table}

It is worth noting that: I) the threshold corrections of Table \ref{tab:ChiC(3P)-splittings} are larger than those of $\chi_{\rm b}(3P)$s, but smaller than those of $\chi_{\rm c}(2P)$ states; see \cite[Table 1]{Ferretti:2018tco}.
In light of this, we expect the $\chi_{\rm c}(3P)$ states to be dominated by the $c \bar c$ core component; II) at present, the only decay mode of the $X(4274)$ which has been observed experimentally is that into $J/\psi \phi$. This may be compatible with the interpretation of the $X(4274)$ as a multiquark state with non-zero hidden-charm hidden-strange components. However, as discussed in Sec. \ref{Open-charm strong decays}, several properties of the $X(4274)$ (e.g. its total decay width) are compatible with those of a $\chi_{\rm c1}(3P)$ state. 

In conclusion, the present results indicate that the $X(4274)$'s wave function should be dominated by the  $\chi_{\rm c1}(3P)$ component.
More informations on the other members of the $\chi_{\rm c}(3P)$ multiplet, as well as a more rigorous analysis of the $X(4274)$'s decay modes, are needed to provide further indications on the quark structure of the previous resonance.

\section{$X(4274)$: other interpretations}
\label{other interpretations}
As pointed out in Ref.~\cite{Ali:2017jda}, molecular states cannot account for the $1^+$ nature of the $X(4274)$.
A possible interpretation of the $X(4274)$ is that of a $s\bar s c \bar c$ compact tetraquark state.
The spectrum of strange and nonstrange hidden-charm compact tetraquark states was computed in Ref.~\cite{Anwar:2018sol} within a relativized diquark-antidiquark model.
There, the authors could provide tetraquark assignments to 13 suspected $XYZ$ exotics, including the $Z_{\rm c}(3900)$, $X(4500)$ and $X(4700)$; however, they could not accommodate the $X(4274)$ within the tetraquark picture.
A similar investigation on $s\bar s c \bar c$ compact tetraquarks was conducted within the relativized quark model \cite{Lu:2016cwr}. There, the authors discussed possible assignments to the $X(4140)$, $X(4500)$ and $X(4700)$, but they could not accommodate the $X(4274)$ within a $s\bar s c \bar c$ compact tetraquark description \cite{Lu:2016cwr}.
In Ref.~\cite{Agaev:2017foq}, the authors made use of QCD sum rules to study the properties of the $X(4140)$ and $X(4274)$.
They interpreted the $X(4140)$ as a $1^{++}$ diquark-antidiquark compact tetraquark in the $\bar {\bf 3}_{\rm c} {\bf 3}_{\rm c}$ color configuration, while the $X(4274)$ was described as a diquark-antidiquark bound state with a ${\bf 6}_{\rm c} \bar {\bf 6}_{\rm c}$ color wave function.
Finally, in Ref.~\cite{Maiani:2016wlq} it was suggested that the $X(4140)$, $X(4274)$, $X(4500)$ and $X(4700)$ could be accommodated within two tetraquark multiplets, with the $X(4274)$ characterized by $0^{++}$ or $2^{++}$ quantum numbers.

In Ref.~\cite{Ortega:2016hde}, the authors investigated possible assignments for the four $J/\psi \phi$ structures, reported by LHCb, CMS, D0 and BaBar \cite{Aaij:2012pz,Chatrchyan:2013dma,Abazov:2013xda,Lees:2014lra}, in a coupled channel scheme by using a nonrelativistic constituent quark model \cite{Vijande:2004he}. 
In particular, they showed that the $X(4274)$, $X(4500)$ and $X(4700)$ can be described as conventional $3^3P_1$, $4^3P_0$, and $5^3P_0$ charmonium states, respectively.
The same interpretation for the $X(4274)$ was proposed in Ref.~\cite{Kher:2018wtv}.
In a study of heavy quarkonium hybrids based on the strong coupling regime of pNRQCD \cite{Oncala:2017hop}, the authors found out that the $X(4274)$ is compatible with a $\chi_{\rm c1}(3P)$ state, which may be affected by the $D_{\rm s}^{*+} D_{\rm s}^{*-}$ threshold.

In Ref.~\cite{He:2016pfa}, an interpretation of the $X(4274)$ as a $P$-wave $D_{\rm s} \bar D_{\rm s0}(2317)$ molecular state in a quasi-potential Bethe-Salpeter equation approach was proposed. If the previous state is a hadronic molecule, an $S$-wave $D_{\rm s} \bar D_{\rm s0}(2317)$ bound state below the $J/\psi \phi$ threshold should also exist.
Finally, in Ref.~\cite{Panteleeva:2018ijz} the authors suggested to assign the $X(4274)$ to a $\psi(2S) \phi$ $S$-wave hadrocharmonium configuration.

\begin{table*}
\centering
\footnotesize
\begin{ruledtabular}
\begin{tabular}{cccc|cccc} 
Process & $P_{c\bar{c}}(A)$ & $P_{c\bar{c}}(B)$ & $\Gamma_{\rm UQM}$ [keV] & Process & $P_{c\bar{c}}(A)$ & $P_{c\bar{c}}(B)$ & $\Gamma_{\rm UQM}$ [keV] \\  
\hline                     
$\psi(2S) \rightarrow \chi_{\rm c2}(1P) + \gamma$ & 0.660 & 0.660 & 16 & $\psi(2S) \rightarrow \chi_{\rm c1}(1P) + \gamma$ & 0.660 & 0.670 & 24 \\
$\psi(2S) \rightarrow \chi_{\rm c0}(1P)  + \gamma$ & 0.660 & 0.700 & 29 & $\eta_{\rm c}(2S) \rightarrow h_{\rm c}(1P)  + \gamma$ & 0.720 & 0.670 & 24 \\
$\psi(3S) \rightarrow \chi_{\rm c2}(2P) + \gamma$ & 0.940 & 0.610 & 8 & $\psi(3S) \rightarrow \chi_{\rm c1}(2P)  + \gamma$ & 0.940 & 0.790 & 29 \\
$\psi(3S) \rightarrow \chi_{\rm c0}(2P)  + \gamma$ & 0.940 & 0.770 & 39 & $\eta_{\rm c}(3S) \rightarrow h_{\rm c}(2P)  + \gamma$ & 0.930 & 0.760 & 75 \\
$\psi(3S) \rightarrow \chi_{\rm c2}(1P)  + \gamma$ & 0.940 & 0.660 & 0 & $\psi(3S) \rightarrow \chi_{\rm c1}(1P)  + \gamma$ & 0.940 & 0.670 & 0 \\
$\psi(3S) \rightarrow \chi_{\rm c0}(1P)  + \gamma$ & 0.940 & 0.700 & 0 & $\eta_c(3S) \rightarrow h_{\rm c}(1P)  + \gamma$ & 0.930 & 0.670 & 7 \\
$\psi(4S) \rightarrow \chi_{\rm c2}(3P)  + \gamma$ & 1.000 & 1.000 & 68 & $\psi(4S) \rightarrow \chi_{\rm c1}(3P)  + \gamma$ & 1.000 & 1.000 & 126 \\
$\psi(4S) \rightarrow \chi_{\rm c0}(3P)  + \gamma$ & 1.000 & 0.960 & 125 & $\eta_c(4S) \rightarrow h_{\rm c}(3P) + \gamma$ & 1.000 & 1.000 & 158 \\
$\psi(4S) \rightarrow \chi_{\rm c2}(2P)  + \gamma$ & 1.000 & 0.610 & 0 & $\psi(4S) \rightarrow \chi_{\rm c1}(2P)  + \gamma$ & 1.000 & 0.790 & 0 \\
$\psi(4S) \rightarrow \chi_{\rm c0}(2P)  + \gamma$ & 1.000 & 0.770 & 0 & $\eta_{\rm c}(4S) \rightarrow h_{\rm c}(2P)  + \gamma$ & 1.000 & 0.760 & 7 \\
$\psi(4S) \rightarrow \chi_{\rm c2}(1P)  + \gamma$ & 1.000 & 0.660 & 0 & $\psi(4S) \rightarrow \chi_{\rm c1}(1P)  + \gamma$ & 1.000 & 0.670 & 0 \\
$\psi(4S) \rightarrow \chi_{\rm c0}(1P)  + \gamma$ & 1.000 & 0.700 & 0 & $\eta_{\rm c}(4S) \rightarrow h_{\rm c}(1P)  + \gamma$ & 1.000 & 0.670 & 3 \\
$\chi_{\rm c2}(1P) \rightarrow J/\psi + \gamma$ & 0.660 & 0.770 & 216 & $\chi_{\rm c1}(1P) \rightarrow J/\psi + \gamma$ & 0.670 & 0.770 & 166 \\
$\chi_{\rm c0}(1P) \rightarrow J/\psi + \gamma$ & 0.700 & 0.770 & 84 & $h_{\rm c}(1P) \rightarrow \eta_{\rm c} + \gamma$ & 0.670 & 0.800 & 267 \\
$\chi_{\rm c2}(2P) \rightarrow \psi(2S)  + \gamma$ & 0.610 & 0.660 & 124 & $\chi_{\rm c1}(2P) \rightarrow \psi(2S)  + \gamma$ & 0.790 & 0.660 & 97 \\
$\chi_{\rm c0}(2P) \rightarrow \psi(2S)  + \gamma$ & 0.770 & 0.660 & 33 & $h_{\rm c}(2P) \rightarrow \eta_c(2S)  + \gamma$ & 0.760 & 0.720 & 151 \\
$\chi_{\rm c2}(2P) \rightarrow J/\psi + \gamma$ & 0.610 & 0.770 & 37 & $\chi_{\rm c1}(2P) \rightarrow J/\psi + \gamma$ & 0.790 & 0.770 & 42 \\
$\chi_{\rm c0}(2P) \rightarrow J/\psi + \gamma$ & 0.770 & 0.770 & 32 & $h_{\rm c}(2P) \rightarrow \eta_c + \gamma$ & 0.760 & 0.800 & 86 \\
$\chi_{\rm c2}(2P) \rightarrow \psi_3(1^3{\rm D}_3) + \gamma$ & 0.610 & 0.660 & 35 & $\chi_{\rm c2}(2P) \rightarrow \psi_2(1^3{\rm D}_2) + \gamma $ & 0.610 & 0.570 & 6 \\
$\chi_{\rm c2}(2P) \rightarrow \psi(1^3{\rm D}_1) + \gamma$ & 0.610 & 0.680 & 1 & $\chi_{\rm c1}(2P) \rightarrow \psi_2(1^3{\rm D}_2)  + \gamma$ & 0.790 & 0.570 & 16 \\
$\chi_{\rm c1}(2P) \rightarrow \psi(1^3{\rm D}_1) + \gamma$ & 0.790 & 0.680 & 12 & $\chi_{\rm c0}(2P) \rightarrow \psi(1^3{\rm D}_1)  + \gamma$ & 0.770 & 0.680 & 7 \\   
$h_{\rm c}(2P) \rightarrow \eta_{\rm c2}(1^1{\rm D}_2)  + \gamma$ & 0.760 & 0.590 & 27 & $\psi_3(1^3{\rm D}_3) \rightarrow \chi_{\rm c2}(1P)  + \gamma$ & 0.660 & 0.660 & 119 \\   
$\psi_2(1^3{\rm D}_2) \rightarrow \chi_{\rm c2}(1P)  + \gamma$ & 0.570 & 0.660 & 24 & $\psi_2(1^3{\rm D}_2) \rightarrow \chi_{\rm c1}(1P)  + \gamma$ & 0.570 & 0.670 & 120 \\
$\psi_1(1^3{\rm D}_1) \rightarrow \chi_{\rm c2}(1P)  + \gamma$ & 0.680 & 0.660 & 2 & $\psi_1(1^3{\rm D}_1) \rightarrow \chi_{\rm c1}(1P)  + \gamma$ & 0.680 & 0.670 & 57 \\
$\psi_1(1^3{\rm D}_1) \rightarrow \chi_{\rm c0}(1P)  + \gamma$ & 0.680 & 0.700 & 193 & $\eta_{\rm c2}(1^1{\rm D}_2) \rightarrow h_c(1P)  + \gamma$ & 0.590 & 0.670 & 136 \\
$\psi_3(2^3{\rm D}_3) \rightarrow \chi_{\rm c2}(2P)  + \gamma$ & 0.960 & 0.610 & 140 & $\psi_2(2^3{\rm D}_2) \rightarrow \chi_{\rm c2}(2P)  + \gamma$ & 0.970 & 0.610 & 31 \\
$\psi_2(2^3{\rm D}_2) \rightarrow \chi_{\rm c1}(2P)  + \gamma$ & 0.970 & 0.790 & 230 & $\psi_1(2^3{\rm D}_1) \rightarrow \chi_{\rm c2}(2P)  + \gamma$ & 0.960 & 0.610 & 3 \\
$\psi_1(2^3{\rm D}_1) \rightarrow \chi_{\rm c1}(2P)  + \gamma$ & 0.960 & 0.790 & 128 & $\psi_1(2^3{\rm D}_1) \rightarrow \chi_{\rm c0}(2P)  + \gamma$ & 0.960 & 0.770 & 360 \\
$\eta_{\rm c2}(2^1{\rm D}_2) \rightarrow h_{\rm c}(2P)  + \gamma$ & 0.970 & 0.760 & 249 & $\psi_3(2^3{\rm D}_3) \rightarrow \chi_{\rm c2}(1P)  + \gamma$ & 0.960 & 0.660 & 18 \\
$\psi_2(2^3{\rm D}_2) \rightarrow \chi_{\rm c2}(1P)  + \gamma$ & 0.970 & 0.660 & 4 & $\psi_2(2^3{\rm D}_2) \rightarrow \chi_{\rm c1}(1P)  + \gamma$ & 0.970 & 0.670 & 17 \\
$\psi_1(2^3{\rm D}_1) \rightarrow \chi_{\rm c2}(1P)  + \gamma$ & 0.960 & 0.660 & 0 &$\psi_1(2^3{\rm D}_1) \rightarrow \chi_{\rm c1}(1P)  + \gamma$ & 0.960 & 0.670 & 9 \\
$\psi_1(2^3{\rm D}_1) \rightarrow \chi_{\rm c0}(1P)  + \gamma$ & 0.960 & 0.700 & 18 & $\eta_{\rm c2}(2^1{\rm D}_2) \rightarrow h_{\rm c}(1P)  + \gamma$ & 0.970 & 0.670 & 26 \\
$\psi_3(2^3{\rm D}_3) \rightarrow \chi_{\rm c4}(1^3{\rm F}_4)  + \gamma$ & 0.960 & 0.950 & 60 & $\psi_3(2^3{\rm D}_3) \rightarrow \chi_{\rm c3}(1^3{\rm F}_3)  + \gamma$ & 0.960 & 0.930 & 4 \\
$\psi_3(2^3{\rm D}_3) \rightarrow \chi_{\rm c2}(1^3{\rm F}_2)  + \gamma$ & 0.960 & 0.940 & 0 & $\psi_2(2^3{\rm D}_2) \rightarrow \chi_{\rm c3}(1^3{\rm F}_3)  + \gamma$ & 0.970 & 0.930 & 40 \\
$\psi_2(2^3{\rm D}_2) \rightarrow \chi_{\rm c2}(1^3{\rm F}_2)  + \gamma$ & 0.970 & 0.940 & 5 & $\psi_1(2^3{\rm D}_1) \rightarrow \chi_{\rm c2}(1^3{\rm F}_2)  + \gamma$ & 0.960 & 0.940 & 46 \\
$\eta_{\rm c2}(2^1{\rm D}_2)  \rightarrow h_{\rm c3}(1^1{\rm F}_3)  + \gamma$ & 0.970 & 0.930 & 49 & $\chi_{\rm c4}(1^3{\rm F}_4) \rightarrow \psi_3(1^3{\rm D}_3)  + \gamma$ & 0.950 & 0.660 & 208 \\
$\chi_{\rm c3}(1^3{\rm F}_3) \rightarrow \psi_3(1^3{\rm D}_3)  + \gamma$ & 0.930 & 0.660 & 25 & $\chi_{\rm c3}(1^3{\rm F}_3) \rightarrow \psi_2(1^3{\rm D}_2)  + \gamma$ & 0.930 & 0.570 & 190 \\
$\chi_{\rm c2}(1^3{\rm F}_2) \rightarrow \psi_3(1^3{\rm D}_3)  + \gamma$ & 0.940 & 0.660 & 1 & $\chi_{\rm c2}(1^3{\rm F}_2) \rightarrow \psi_2(1^3{\rm D}_2)  + \gamma$ & 0.940 & 0.570 & 34 \\
$\chi_{\rm c2}(1^3{\rm F}_2) \rightarrow \psi_1(1^3{\rm D}_1)  + \gamma$ & 0.940 & 0.680 & 304 & $h_{\rm c3}(1^1{\rm F}_3) \rightarrow \eta_{\rm c2}(1^1{\rm D}_2)  + \gamma$ & 0.930 & 0.590 & 215 \\
$\chi_{\rm c4}(2^3{\rm F}_4) \rightarrow \psi_3(2^3{\rm D}_3)  + \gamma$ & 1.000 & 0.960 & 296 & $\chi_{\rm c3}(2^3{\rm F}_3) \rightarrow \psi_3(2^3{\rm D}_3)  + \gamma$ & 1.000 & 0.960 & 35 \\
$\chi_{\rm c3}(2^3{\rm F}_3) \rightarrow \psi_2(2^3{\rm D}_2)  + \gamma$ & 1.000 & 0.970 & 324 & $\chi_{\rm c2}(2^3{\rm F}_2) \rightarrow \psi_3(2^3{\rm D}_3)  + \gamma$ & 1.000 & 0.960 & 1 \\
$\chi_{\rm c2}(2^3{\rm F}_2) \rightarrow \psi_2(2^3{\rm D}_2)  + \gamma$ & 1.000 & 0.970 & 56 & $\chi_{\rm c2}(2^3{\rm F}_2) \rightarrow \psi_1(2^3{\rm D}_1)  + \gamma$ & 1.000 & 0.960 & 294 \\
$h_{\rm c3}(2^1{\rm F}_3) \rightarrow \eta_{\rm c2}(2^1{\rm D}_2)  + \gamma$ & 1.000 & 0.970 & 351 & $\chi_{\rm c4}(2^3{\rm F}_4)  \rightarrow \psi_3(1^3{\rm D}_3)  + \gamma$ & 1.000 & 0.660 & 13 \\
$\chi_{\rm c3}(2^3{\rm F}_3) \rightarrow \psi_3(1^3{\rm D}_3)  + \gamma$ & 1.000 & 0.660 & 1 & $\chi_{\rm c3}(2^3{\rm F}_3) \rightarrow \psi_2(1^3{\rm D}_2)  + \gamma$ & 1.000 & 0.570 & 11 \\
$\chi_{\rm c2}(2^3{\rm F}_2) \rightarrow \psi_3(1^3{\rm D}_3)  + \gamma$ & 1.000 & 0.660 & 0 & $\chi_{\rm c2}(2^3{\rm F}_2) \rightarrow \psi_2(1^3{\rm D}_2)  + \gamma$ & 1.000 & 0.570 & 2 \\
$\chi_{\rm c2}(2^3{\rm F}_2) \rightarrow \psi_1(1^3{\rm D}_1)  + \gamma$ & 1.000 & 0.680 & 14 & $h_{\rm c3}(2^1{\rm F}_3) \rightarrow \eta_{\rm c2}(1^1{\rm D}_2)  + \gamma$ & 1.000 & 0.590 & 13 \\
$\chi_{\rm c4}(2^3{\rm F}_4) \rightarrow \psi_5(1^3{\rm G}_5)  + \gamma$ & 1.000 & 0.980 & 53 & $\chi_{\rm c4}(2^3{\rm F}_4) \rightarrow \psi_4(1^3{\rm G}_4)  + \gamma$ & 1.000 & 1.000 & 2 \\
$\chi_{\rm c4}(2^3{\rm F}_4) \rightarrow \psi_3(1^3{\rm G}_3)  + \gamma$ & 1.000 & 1.000 & 0 & $\chi_{\rm c3}(2^3{\rm F}_3) \rightarrow \psi_4(1^3{\rm G}_4)  + \gamma$ & 1.000 & 1.000 & 43 \\
$\chi_{\rm c3}(2^3{\rm F}_3) \rightarrow \psi_3(1^3{\rm G}_3)  + \gamma$ & 1.000 & 1.000 & 2 & $\chi_{\rm c2}(2^3{\rm F}_2) \rightarrow \psi_3(1^3{\rm G}_3)  + \gamma$ & 1.000 & 1.000 & 36 \\
$h_{\rm c3}(2^1{\rm F}_3) \rightarrow \eta_{\rm c4}(1^1{\rm G}_4)  + \gamma$ & 1.000 & 1.000 & 47 & $\psi_5(1^3{\rm G}_5) \rightarrow \chi_{\rm c4}(1^3{\rm F}_4)  + \gamma$ & 0.980 & 0.950 & 345 \\
$\psi_4(1^3{\rm G}_4) \rightarrow \chi_{\rm c4}(1^3{\rm F}_4)  + \gamma$ & 1.000 & 0.950 & 27 & $\psi_4(1^3{\rm G}_4) \rightarrow \chi_{\rm c3}(1^3{\rm F}_3)  + \gamma$ & 1.000 & 0.930 & 356 \\
$\psi_3(1^3{\rm G}_3) \rightarrow \chi_{\rm c4}(1^3{\rm F}_4)  + \gamma$ & 1.000 & 0.950 & 1 & $\psi_3(1^3{\rm G}_3) \rightarrow \chi_{\rm c3}(1^3{\rm F}_3)  + \gamma$ & 1.000 & 0.930 & 35 \\
$\psi_3(1^3{\rm G}_3) \rightarrow \chi_{\rm c2}(1^3{\rm F}_2)  + \gamma$ & 1.000 & 0.940 & 401 & $\eta_{\rm c4}(1^1{\rm G}_4) \rightarrow h_{\rm c3}(1^1{\rm F}_3)  + \gamma$ & 1.000 & 0.930 & 380 \\
\end{tabular}
\end{ruledtabular}
\caption{As Table \ref{tab:ChiC(3P)-E1}, but for the radiative transitions of different charmonia. Our UQM predictions, $\Gamma_{E1, \rm UQM}$, are computed according to Eq. (\ref{eqn:E1-UQM}). The large values of the $P_{c\bar{c}}$ probabilities for several charmonia are due to the conservative renormalization prescription used in our UQM calculation of the radiative decays. See the explanation in Sec.\,\ref{E1 charmonium} and also in Ref. \cite{Anwar:2018yqm}.}
\label{tab:others-E1}  
\end{table*}

\section{Conclusion}
We studied the quark structure, the spectrum and the strong open-charm and radiative decay modes of the $X(4274)$ and $\chi_{\rm c}(3P)$ states within an UQM-based coupled-channel model \cite{Ferretti:2018tco} and the quark model formalism \cite{Micu,LeYaouanc,Roberts:1992,Eichten:1978tg,Gupta,Kwong:1988ae}. 

The present coupled-channel model was previously used to study the properties and quark structure of the $\chi_{\rm c}(2P)$ and $\chi_{\rm b}(3P)$ multiplets \cite{Ferretti:2018tco}. There, a prescription to ``renormalize'' the UQM results for the self-energy/threshold corrections made it possible to distinguish between quarkonia, the $\chi_{\rm b}(3P)$, and quarkonium-like states with significant meson-meson components in their wave functions, the $\chi_{\rm c}(2P)$s.

According to our new results, the $X(4274)$ can be described as a $\chi_{\rm c1}(3P)$ state. The other members of the $\chi_{\rm c}(3P)$ multiplet can be interpreted as $3P$ charmonium cores plus small to medium-sized open-charm meson-meson components.

A comparison between theoretical results for the radiative transitions of $\chi_{\rm c}(3P)$s (including ours and, for example, those from Ref. \cite{Barnes:2005pb}) and the forthcoming experimental data may provide exploratory pathways to search for still unobserved $3P$ charmonia.
Hence, we suggest the experimentalists to focus on the study of the $\chi_{\rm c}(3P) \rightarrow \psi(nS)+\gamma$ decay modes, and especially on the $\psi(2S,3S) + \gamma$ transitions.

In conclusion, we hope that this study might be helpful to fulfill a better understanding of higher $P$-wave charmonia.
More precise conclusions regarding the quark structure of the $\chi_{\rm c}(3P)$ states will necessarily require more experimental informations on the properties of the still unobserved $h_{\rm c}(3P)$, $\chi_{\rm c0}(3P)$ and $\chi_{\rm c2}(3P)$.

\begin{acknowledgments}
We are grateful to Ulf-G.\,Mei{\ss}ner for a careful reading of the manuscript,
and to Bing-Song Zou for mentoring on the formalism of this manuscript.
  J.\,Ferretti acknowledges financial support from the US Department of Energy, Grant No. DE-FG-02-91ER-40608, and the Academy of Finland, Project no. 320062. Y.\,Lu and M.\,Naeem\,Anwar are supported by the DFG (Grant No. TRR110) and the NSFC (Grant No. 11621131001) through the funds provided to the Sino-German CRC 110 ``Symmetries and the Emergence of Structure in QCD''. M. N. Anwar acknowledges partial support from the Munich Institute for Astro- and Particle Physics (MIAPP) which is funded by the DFG under Germany's Excellence Strategy$-$EXC-2094$-$390783311.
\end{acknowledgments}

\begin{table*}[htbp]
\centering
\begin{ruledtabular}
\begin{tabular}{ccccc} 
Process & $P_{c\bar{c}}(A)$ & $P_{c\bar{c}}(B)$ & $\Gamma_{\rm UQM}$ [keV] & $\Gamma_{\rm exp}$ [keV]  \\  
\hline                     
$\psi(2S) \rightarrow \chi_{\rm c2}(1P) + \gamma$ & 0.660 & 0.660 & 16  & $27.9\pm0.6$ \\
$\psi(2S) \rightarrow \chi_{\rm c1}(1P) + \gamma$ & 0.660 & 0.670 & 24  & $28.7\pm0.7$ \\
$\psi(2S) \rightarrow \chi_{\rm c0}(1P)  + \gamma$ & 0.660 & 0.700 & 29 & $28.8\pm0.6$ \\
$\chi_{\rm c2}(1P) \rightarrow J/\psi + \gamma$ & 0.660 & 0.770 & 216 & $374.3\pm10$ \\
$\chi_{\rm c1}(1P) \rightarrow J/\psi + \gamma$ & 0.670 & 0.770 & 166 & $288\pm8.4$ \\
$\chi_{\rm c0}(1P) \rightarrow J/\psi + \gamma$ & 0.700 & 0.770 & 84 & $151.2\pm5.4$ \\
$h_{\rm c}(1P) \rightarrow \eta_{\rm c} (1S)+ \gamma$ & 0.670 & 0.800 & 267 & $357\pm42$ \\
$\psi_1(1^3D_1) \rightarrow \chi_{\rm c2}(1P)  + \gamma$ & 0.680 & 0.660 & 2 & $<17.4$ \\
$\psi_1(1^3D_1) \rightarrow \chi_{\rm c1}(1P)  + \gamma$ & 0.680 & 0.670 & 57 & $67.7\pm6$ \\
$\psi_1(1^3D_1) \rightarrow \chi_{\rm c0}(1P)  + \gamma$ & 0.680 & 0.700 & 193 &$187.7\pm16$ \\
\end{tabular}
\end{ruledtabular}
\caption{Our UQM predictions for $E1$ radiative decay widths of lower charmonia are compared to the available experimental results \cite{Tanabashi:2018oca}.}
\label{tab:ChiC(3P)-E1-M1-exp}  
\end{table*}

\begin{appendix}

\section{$E1$ radiative transitions of charmonium states}
\label{appendix A}
In Table \ref{tab:others-E1}, we enlist our UQM results for the $E1$ radiative transition widths of higher-lying charmonia, including $2S$, $3S$, $1P$ and $2P$ resonances. The widths are calculated as explained in Sec. \ref{E1 charmonium}. In Table \ref{tab:ChiC(3P)-E1-M1-exp}, we compare our UQM predictions to the available experimental data \cite{Tanabashi:2018oca}. Our results can also be compared to the QM predictions of Ref. \cite{Barnes:2005pb}.

It is worth noting that our predictions are in good accordance with the existing experimental results \cite{Tanabashi:2018oca}.
This is a further indication of the importance of the radiative transitions in the study of the properties of both the well-established and still unobserved heavy quarkonium resonances.

\end{appendix}

\end{document}